\documentclass[10pt]{article}
\usepackage{graphicx}
\usepackage{amsmath}
\usepackage{amssymb}
\usepackage{caption2}
\begin{document}
\begin{center}
{\large {\bf \sc{  Analysis of the  mass and width of the $X(4140)$ as axialvector tetraquark state
  }}} \\[2mm]
Zhi-Gang  Wang \footnote{E-mail: zgwang@aliyun.com.  }, Zun-Yan Di   \\
 Department of Physics, North China Electric Power University, Baoding 071003, P. R. China
\end{center}

\begin{abstract}
In this article, we construct  both the $[sc]_T[\bar{s}\bar{c}]_A+[sc]_A[\bar{s}\bar{c}]_T$ type and $[sc]_T[\bar{s}\bar{c}]_V-[sc]_V[\bar{s}\bar{c}]_T$ type axialvector currents with $J^{PC}=1^{++}$ to study the mass of the $X(4140)$ with the QCD sum rules. The predicted masses support assigning the $X(4140)$ to be the $[sc]_T[\bar{s}\bar{c}]_V-[sc]_V[\bar{s}\bar{c}]_T$ type axialvector tetraquark state. Then we study the hadronic coupling constant $g_{XJ/\psi\phi}$ with the QCD sum rules based on  solid quark-hadron duality, and obtain the decay width $\Gamma(X(4140)\to J/\psi \phi)=86.9\pm22.6\,{\rm{MeV}}$, which is in excellent agreement with the experimental data $83\pm 21^{+21}_{-14} {\mbox{ MeV}}$ from the LHCb collaboration.
\end{abstract}



\section{Introduction}

In 2009,  the CDF collaboration observed the $X(4140)$ for the first time  in the $J/\psi\phi$  mass spectrum in the exclusive $B^+ \to J/\psi\,\phi K^+$ decays in $p\bar{p}$ collisions with a statistical
significance  more than $3.8 \sigma$  \cite{CDF0903}. Then the $X(4140)$ was confirmed by CDF, CMS, D0, LHCb collaborations  \cite{CDF0903,CDF1101,CMS1309,D0-1309,D0-1508,LHCb-16061,LHCb-16062}.
The LHCb collaboration performed the first full amplitude analysis of the decays $B^+\to J/\psi \phi K^+$ and  confirmed the two old particles $X(4140)$ and $X(4274)$ in the $J/\psi \phi$  mass spectrum  with statistical significances $8.4\sigma$ and $6.0\sigma$, respectively,  and determined the    spin-parity-change-conjugation  to be $J^{PC} =1^{++}$ with statistical significances $5.7\sigma$ and $5.8\sigma$, respectively \cite{LHCb-16061,LHCb-16062}. In Table 1, we present the mass, width, $J^{PC}$ of the  $X(4140)$ from the different experiments.  Although the width  from the LHCb collaboration \cite{LHCb-16061,LHCb-16062} differs  from other measurements greatly, the masses from different experiments are consistent with each other. The $D_s^*\bar{D}_s^*$ threshold is $4224.4\,\rm{MeV}$ from the Particle Data Group \cite{PDG}, which leads to the possible molecule assignment for the $X(4140)$.   The $X(4140)$ was observed in the final state $J/\psi\phi$, its  $J^{PC}=0^{++}$, $1^{++}$, $2^{++}$ for the S-wave couplings, and $0^{-+}$, $1^{-+}$, $2^{-+}$, $3^{-+}$ for the P-wave couplings. The most popular current to interpolate the $D_s^*$ meson is $J_\alpha(x)=\bar{s}(x)\gamma_\alpha c(x)$, the most popular current to interpolate the $D_s^*\bar{D}_s^*$ molecular states is $J_{\alpha\beta}(x)=\bar{s}(x)\gamma_\alpha c(x)\bar{c}(x)\gamma_\beta s(x)$. We can study the $J^{PC}=0^{++}$, $1^{-+}$, $2^{++}$, $1^{+-}$, $1^{--}$  $D_s^*\bar{D}_s^*$ molecular states with the QCD sum rules by using the suitable projectors.
The LHCb collaboration determined the quantum numbers of the $X(4140)$ to be $J^{PC}=1^{++}$, which rules  out the $0^{++}$ or $2^{++}$ $D_s^*\bar{D}_s^*$ molecule assignment, but does not rule  out the existence of the $0^{++}$ or $2^{++}$ $D_s^*\bar{D}_s^*$ molecular states.

 The  possible assignments for the $X(4140)$ are   tetraquark state \cite{X4140-tetraquark-Stancu,Wang1502-Y4140,X4140-tetraquark-Lebed,Maiani-X4140,X4140-tetraquark}, hybrid state \cite{X4140-hybrid, X4140-hybrid-Wang0903} or  rescattering effect \cite{X4140-rescat}, etc.
In Ref.\cite{X4140-tetraquark-Stancu}, F. Stancu calculates  the mass spectrum of the $c\bar{c}s\bar{s}$ tetraquark states   via   a simple quark model with chromomagnetic interaction, and obtain two lowest masses $4195\,\rm{MeV}$ and $4356\,\rm{MeV}$  with $J^{PC}=1^{++}$. The value $4195\,\rm{MeV}$ is consistent with the LHCb data $4146.5 \pm 4.5 ^{+4.6}_{-2.8} \mbox{ MeV}$ \cite{LHCb-16061,LHCb-16062}. In the simple chromomagnetic interaction model, there are no correlated quarks or diquarks \cite{X4140-tetraquark-Stancu}. In Ref.\cite{X4140-tetraquark-Lebed}, R. F. Lebed  and A. D. Polosa assign the $X(4140)$ to  be the $J^{PC}=1^{++}$ diquark-antidiquark state $[cs]_{A}[\bar{c}\bar{s}]_{S}+[cs]_{S}[\bar{c}\bar{s}]_{A}$  based on the effective  Hamiltonian with the spin-spin and spin-orbit  interactions, then in Ref.\cite{Maiani-X4140}, L. Maiani, A. D. Polosa and V. Riquer take the mass of the $X(4140)$ as input parameter, and obtain the mass spectrum of the $c\bar{c}s\bar{s}$ tetraquark states with positive parity, however, they observe that there is no room for the $X(4274)$, and suggest the $X(4274)$
 corresponds to two, almost degenerate, unresolved lines with $J^{PC}=0^{++}$ and $2^{++}$.

In the QCD sum rules, we usually take the diquarks (or correlations) and antidiquarks (or correlations) as the basic constituents to construct the interpolating currents, the predictions can be compared to that based on the diquark-antidiquark model directly \cite{X4140-tetraquark-Lebed,Maiani-X4140}.  In the quantum field theory, the diquark operators (or diquarks) $\varepsilon^{ijk}q^{T}_j C\Gamma q^{\prime}_k$ have  five  structures  in Dirac spinor space, where the $i$, $j$ and $k$ are color indexes,  $C\Gamma=C\gamma_5$, $C$, $C\gamma_\mu \gamma_5$,  $C\gamma_\mu $ and $C\sigma_{\mu\nu}$ for the scalar ($S$), pseudoscalar ($P$), vector ($V$), axialvector ($A$)  and  tensor ($T$) diquarks, respectively.  The $C\gamma_5$ and $C\gamma_\mu$ diquark states have the spin-parity $J^P=0^+$ and $1^+$, respectively,  the $C$ and $C\gamma_\mu\gamma_5$ diquark states have the spin-parity $J^P=0^-$ and $1^-$, respectively, the $C\sigma_{\mu\nu}$ and $C\sigma_{\mu\nu}\gamma_5$ diquark states (or operators) have both the $J^P=1^+$ and $1^-$ components. The relevant diquark-antidiquark type scalar, axialvector and tensor tetraquark   states $sc\bar{s}\bar{c}$ have been studied with the QCD sum rules \cite{Wang1502-Y4140,Wang1606-Y3915C,Wang1607-Y3915A,Wang1607-Y4140,ChenZhu-2011,Chen1606,Azizi1703,ChenZhu1706}, see Table 2.

In the QCD sum rules for the hidden-charm (or hidden-bottom) tetraquark states and molecular states, the integrals
 \begin{eqnarray}
 \int_{4m_Q^2(\mu)}^{s_0} ds \rho_{QCD}(s,\mu)\exp\left(-\frac{s}{T^2} \right)\, ,
 \end{eqnarray}
are sensitive to the energy scales $\mu$, where the $\rho_{QCD}(s,\mu)$ are  the QCD spectral densities, the $T^2$ are the Borel parameters, the $s_0$ are the continuum thresholds parameters, the predicted masses depend heavily on the energy scales $\mu$.
In Refs.\cite{Wang-tetra-formula,WangHuang-molecule}, we suggest an energy scale formula $\mu=\sqrt{M^2_{X/Y/Z}-(2{\mathbb{M}}_Q)^2}$ with the effective $Q$-quark mass ${\mathbb{M}}_Q$ to determine the ideal energy scales of the QCD spectral densities. The formula enhances the pole contributions remarkably, we obtain the pole contributions as large as $(40-60)\%$ in Refs.\cite{Wang1502-Y4140,Wang1606-Y3915C,Wang1607-Y3915A,Wang1607-Y4140}, otherwise, the pole contributions are about $40\%$ \cite{ChenZhu-2011} or $20\%$ \cite{Azizi1703,ChenZhu1706}. The energy scale formula also works well in the QCD sum rules  for the hidden-charm pentaquark states \cite{Wang1508-EPJC}.

From Table 2, we can see that the $[sc]_S[\bar{s}\bar{c}]_A+[sc]_A[\bar{s}\bar{c}]_S$ type axialvector current cannot reproduce the mass of the $X(4140)$  if the pole dominance criterion is satisfied. If we take energy scale formula $\mu=\sqrt{M^2_{X/Y/Z}-(2{\mathbb{M}}_c)^2}$ and choose the updated value ${\mathbb{M}}_c=1.82\,\rm{GeV}$ \cite{WangEPJC-1601}, we can obtain the optimal energy scales $\mu=1.4\,\rm{GeV}$ and $2.0\,\rm{GeV}$ for the QCD spectral densities in  the QCD sum rules for the  $Z_c(3900)$ and $X(4140)$, respectively. In Ref.\cite{Wang1607-Y4140}, we observe that the mass of the $X(4140)$ can be reproduced at the energy scale $\mu=1.1\,\rm{GeV}$, a too low energy scale. The QCD sum rules do no support assigning the $X(4140)$ to be the $[sc]_S[\bar{s}\bar{c}]_A+[sc]_A[\bar{s}\bar{c}]_S$ type axialvector tetraquark state.

\begin{table}
\begin{center}
\begin{tabular}{|c|c|c|c|c|c|c|c|}\hline\hline

Year   &Mass (MeV)                     &Width (MeV) ~~~                 &$J^{PC}$ &Significance         &Experiment     \\ \hline\hline

2009   &$4143.0\pm 2.9\pm 1.2$         &$11.7^{+8.3}_{-5.0}\pm 3.7$     &         &$3.8\,\sigma$        &CDF \cite{CDF0903}     \\ \hline

2011   &$4143.4^{+ 2.9}_{-3.0}\pm 0.6$ &$15.3^{+10.4}_{-6.1}\pm 2.5$    &         &$5.0\,\sigma$        &CDF \cite{CDF1101}    \\  \hline

2013   &$4148.0\pm 2.4\pm 6.3$         &$28^{+15}_{-11}\pm 19$          &         &$5.0\,\sigma$        &CMS \cite{CMS1309}    \\  \hline

2013   &$4159.0\pm 4.3\pm 6.6$         &$19.9\pm 12.6^{+3.0}_{-8.0}$    &         &$3.1\,\sigma$        &D0 \cite{D0-1309}  \\  \hline

2015   &$4152.5\pm 1.7^{+ 6.2}_{-5.4}$ &$16.3\pm 5.6\pm 11.4$           &         &$4.7\,\sigma$        &D0 \cite{D0-1508} \\ \hline

2016   &$4146.5\pm 4.5^{+ 4.6}_{-2.8}$ &$83\pm 21^{+21}_{-14}$          &$1^{++}$ &$8.4\,\sigma$       &LHCb \cite{LHCb-16061,LHCb-16062}   \\ \hline\hline
 \end{tabular}
\end{center}
\caption{ The  mass, width, $J^{PC}$ of the  $X(4140)$ from the different experiments. }
\end{table}

\begin{table}
\begin{center}
\begin{tabular}{|c|c|c|c|c|c|c|c|}\hline\hline
$J^{PC}$   &Structures                                             &OPE\,(No)    &mass(GeV)      &Assignment        &References   \\ \hline

$0^{++}$   &$[sc]_A[\bar{s}\bar{c}]_A$                             &$10$         &$3.92/4.50$    &$X(3915)/X(4500)$ &\cite{Wang1606-Y3915C}  \\ \hline
$0^{++}$   &$[sc]_V[\bar{s}\bar{c}]_V$                             &$10$         &$4.70$         &$X(4700)$         &\cite{Wang1606-Y3915C}  \\ \hline
$0^{++}$   &$[sc]_A[\bar{s}\bar{c}]_A$                             &$10$         &$3.98$         &$?$               &\cite{Wang1502-Y4140}  \\ \hline

$0^{++}$   &$[sc]_S[\bar{s}\bar{c}]_S$                             &$10$         &$3.89/4.35$    &$X(3915)/\,?$     &\cite{Wang1607-Y3915A}  \\ \hline
$0^{++}$   &$[sc]_P[\bar{s}\bar{c}]_P$                             &$10$         &$5.48$         &$?$               &\cite{Wang1607-Y3915A}  \\ \hline

$1^{++}$   &$[sc]_S[\bar{s}\bar{c}]_A+[sc]_A[\bar{s}\bar{c}]_S$    &$10$         &$3.95$         &$?$               &\cite{Wang1607-Y4140}  \\ \hline
$1^{++}$   &$[sc]_P[\bar{s}\bar{c}]_V+[sc]_V[\bar{s}\bar{c}]_P$    &$10$         &$5.00$         &$?$               &\cite{Wang1607-Y4140}  \\ \hline
$1^{++}$   &$[sc]_S[\bar{s}\bar{c}]_A+[sc]_A[\bar{s}\bar{c}]_S$    &$8(7)$       &$4.07$         &$X(4140)$         &\cite{ChenZhu-2011}  \\ \hline
$1^{++}$   &$[sc]_S[\bar{s}\bar{c}]_A+[sc]_A[\bar{s}\bar{c}]_S$    &$8$          &$4.18$         &$X(4140)$         &\cite{Azizi1703}  \\ \hline

$2^{++}$   &$[sc]_A[\bar{s}\bar{c}]_A$                             &$10$         &$4.13$         &$?\,X(4140)$      &\cite{Wang1502-Y4140}  \\ \hline \hline
\end{tabular}
\end{center}
\caption{ The masses of the $sc\bar{s}\bar{c}$ tetraquark states relevant to the $X(4140)$  from the QCD sum rules, the OPE denotes  truncations of  the operator product expansion up to the vacuum condensates of dimension $n$, the No denotes the vacuum condensates of dimension $n^\prime$ are not included.   }
\end{table}

In this article, we construct the
$[sc]_T[\bar{s}\bar{c}]_A+[sc]_A[\bar{s}\bar{c}]_T$ type and $[sc]_T[\bar{s}\bar{c}]_V-[sc]_V[\bar{s}\bar{c}]_T$  type axialvector currents to study the mass of the $X(4140)$ as the axialvector tetraquark state with the QCD sum rules in details, then study the width of the $X(4140)$ with the QCD sum rules based on the solid quark-hadron duality.

The article is arranged as follows:  we derive the QCD sum rules for
the mass and width of the  $X(4140)$ as axialvector tetraquark state   in section 2 and in section 3 respectively;  section 4 is reserved for our conclusion.

\section{The mass of the $X(4140)$ as the axialvector tetraquark state}
In the following, we write down  the two-point correlation functions $\Pi_{\mu\mu^\prime}(p)$  in the QCD sum rules,
\begin{eqnarray}
\Pi_{\mu\mu^\prime}(p)&=&i\int d^4x e^{ip \cdot x} \langle0|T\left\{J_\mu(x)J_{\mu^\prime}^{\dagger}(0)\right\}|0\rangle \, ,
\end{eqnarray}
where $J_\mu(x)=J^1_\mu(x)$, $J^2_\mu(x)$,
\begin{eqnarray}
J_\mu^1(x)&=&\frac{\varepsilon^{ijk}\varepsilon^{imn}}{\sqrt{2}}\left[s^{Tj}(x)C\sigma_{\mu\nu}\gamma_5 c^k(x)\bar{s}^m(x)\gamma^\nu C \bar{c}^{Tn}(x)+s^{Tj}(x)C\gamma^\nu c^k(x)\bar{s}^m(x)\gamma_5\sigma_{\mu\nu} C \bar{c}^{Tn}(x) \right] \, , \nonumber\\
J_\mu^2(x)&=&\frac{\varepsilon^{ijk}\varepsilon^{imn}}{\sqrt{2}}\left[s^{Tj}(x)C\sigma_{\mu\nu} c^k(x)\bar{s}^m(x)\gamma_5\gamma^\nu C \bar{c}^{Tn}(x)-s^{Tj}(x)C\gamma^\nu \gamma_5c^k(x)\bar{s}^m(x) \sigma_{\mu\nu} C \bar{c}^{Tn}(x) \right] \, , \nonumber
\end{eqnarray}
  the $i$, $j$, $k$, $m$, $n$ are  color indexes.
 Under charge conjugation (parity) transform $\widehat{C}$ ($\widehat{P}$), the currents $J_\mu(x)$ have the properties,
\begin{eqnarray}
\widehat{C}J_{\mu}(x)\widehat{C}^{-1}&=&+ J_\mu(x) \, , \nonumber\\
\widehat{P}J_{\mu}(x)\widehat{P}^{-1}&=&- J^\mu(\tilde{x}) \, ,
\end{eqnarray}
the four vectors $x^\mu=(t,\vec{x})$ and $\tilde{x}^\mu=(t,-\vec{x})$. The currents $J^1_\mu(x)$ and $J^2_\mu(x)$ couple potentially to the
$[sc]_T[\bar{s}\bar{c}]_A+[sc]_V[\bar{s}\bar{c}]_A$ and $[sc]_T[\bar{s}\bar{c}]_V-[sc]_V[\bar{s}\bar{c}]_T$
 axialvector tetraquark states with $J^{PC}=1^{++}$, respectively.
The tensor diquark  operators have the  properties,
\begin{eqnarray}
\widehat{P}\varepsilon^{ijk}s^{Tj}(x)C\sigma_{\mu\nu}\gamma_5c^k(x)\widehat{P}^{-1}&=&\varepsilon^{ijk}s^{Tj}(\tilde{x})C\sigma^{\mu\nu}\gamma_5c^k(\tilde{x}) \, , \nonumber\\
\widehat{P}\varepsilon^{ijk}s^{Tj}(x)C\sigma_{\mu\nu} c^k(x)\widehat{P}^{-1}&=&-\varepsilon^{ijk}s^{Tj}(\tilde{x})C\sigma^{\mu\nu} c^k(\tilde{x}) \, ,
\end{eqnarray}
under parity transform, the tensor diquark operators  couple potentially  to both the $J^P=1^+$ and $1^-$ diquark states. We should project out the $1^+$ or $1^-$ component by multiplying tensor diquark operators by the axialvector antidiquark operator $\varepsilon^{imn}\bar{s}^m(x)\gamma^\nu C \bar{c}^{Tn}(x)$ or vector antidiquark operator $\varepsilon^{imn}\bar{s}^m(x)\gamma_5\gamma^\nu C \bar{c}^{Tn}(x)$.

At the phenomenological side,  we insert  a complete set of intermediate hadronic states with
the same quantum numbers as the current operators $J_\mu(x)$ into the
correlation functions $\Pi_{\mu\mu^\prime}(p)$  to obtain the hadronic representation
\cite{SVZ79,Reinders85}, and isolate the ground state
contributions,
\begin{eqnarray}
\Pi_{\mu\mu^\prime}(p)&=&\frac{\lambda_{X}^2}{M^2_{X}-p^2}\left(-g_{\mu\mu^{\prime} } +\frac{p_\mu p_{\mu^{\prime}}}{p^2}\right) +\cdots  \nonumber\\
&=&\Pi(p^2)\left(-g_{\mu\mu^{\prime} } +\frac{p_\mu p_{\mu^{\prime}}}{p^2}\right) +\cdots \, ,
\end{eqnarray}
where the pole residues  $\lambda_{X}$ are defined by $\langle 0|J_\mu(0)|X(p)\rangle=\lambda_{X}\, \varepsilon_\mu$,
the $\varepsilon_\mu$ are  the polarization vectors of the axialvector tetraquark states $X$.

Now  we briefly outline  the operator product expansion for the correlation functions $\Pi_{\mu\mu^\prime}(p)$.  We contract the quark fields $s$ and $c$ in the correlation functions
$\Pi_{\mu\mu^\prime}(p)$ with Wick theorem, and obtain the results,
\begin{eqnarray}
\Pi^1_{\mu\mu^{\prime}}(p)&=&-\frac{i}{2} \varepsilon^{ijk}\varepsilon^{imn}\varepsilon^{i^{\prime}j^{\prime}k^{\prime}}\varepsilon^{i^{\prime}m^{\prime}n^{\prime}}   \int d^4x e^{ip \cdot x}   \nonumber\\
&&\left\{{\rm Tr}\left[ \sigma_{\mu\nu}\gamma_5 S_c^{kk^{\prime}}(x)\gamma_5\sigma_{\mu^{\prime}\nu^{\prime}} CS^{Tjj^{\prime}}(x)C\right] {\rm Tr}\left[ \gamma^{\nu^\prime} S_c^{n^{\prime}n}(-x)\gamma^\nu CS^{Tm^{\prime}m}(-x)C\right] \right. \nonumber\\
&&+{\rm Tr}\left[ \gamma^\nu S_c^{kk^{\prime}}(x)\gamma_5\sigma_{\mu^{\prime}\nu^{\prime}} CS^{Tjj^{\prime}}(x)C\right] {\rm Tr}\left[ \gamma^{\nu^{\prime}} S_c^{n^{\prime}n}(-x)\gamma_5 \sigma_{\mu\nu}CS^{Tm^{\prime}m}(-x)C\right] \nonumber\\
&&+ {\rm Tr}\left[ \sigma_{\mu\nu}\gamma_5 S_c^{kk^{\prime}}(x)\gamma^{\nu^{\prime}} CS^{Tjj^{\prime}}(x)C\right] {\rm Tr}\left[ \sigma_{\mu^{\prime}\nu^{\prime}}\gamma_5 S_c^{n^{\prime}n}(-x)\gamma^\nu CS^{Tm^{\prime}m}(-x)C\right] \nonumber\\
 &&\left.+ {\rm Tr}\left[ \gamma^{\nu} S_c^{kk^{\prime}}(x)\gamma^{\nu^{\prime}}CS^{Tjj^{\prime}}(x)C\right] {\rm Tr}\left[\sigma_{\mu^{\prime}\nu^{\prime}} \gamma_5 S_c^{n^{\prime}n}(-x)\gamma_5\sigma_{\mu\nu} CS^{Tm^{\prime}m}(-x)C\right] \right\} \, ,
\end{eqnarray}

\begin{eqnarray}
\Pi^2_{\mu\mu^{\prime}}(p)&=&-\frac{i}{2} \varepsilon^{ijk}\varepsilon^{imn}\varepsilon^{i^{\prime}j^{\prime}k^{\prime}}\varepsilon^{i^{\prime}m^{\prime}n^{\prime}}   \int d^4x e^{ip \cdot x}   \nonumber\\
&&\left\{{\rm Tr}\left[ \sigma_{\mu\nu}S_c^{kk^{\prime}}(x)\sigma_{\mu^{\prime}\nu^{\prime}} CS^{Tjj^{\prime}}(x)C\right] {\rm Tr}\left[ \gamma^{\nu^\prime} \gamma_5 S_c^{n^{\prime}n}(-x)\gamma_5 \gamma^\nu CS^{Tm^{\prime}m}(-x)C\right] \right. \nonumber\\
&&+{\rm Tr}\left[ \gamma^\nu\gamma_5  S_c^{kk^{\prime}}(x)\sigma_{\mu^{\prime}\nu^{\prime}} CS^{Tjj^{\prime}}(x)C\right] {\rm Tr}\left[ \gamma^{\nu^{\prime}} \gamma_5 S_c^{n^{\prime}n}(-x) \sigma_{\mu\nu}CS^{Tm^{\prime}m}(-x)C\right] \nonumber\\
&&+ {\rm Tr}\left[ \sigma_{\mu\nu} S_c^{kk^{\prime}}(x)\gamma_5 \gamma^{\nu^{\prime}} CS^{Tjj^{\prime}}(x)C\right] {\rm Tr}\left[ \sigma_{\mu^{\prime}\nu^{\prime}} S_c^{n^{\prime}n}(-x)\gamma_5 \gamma^\nu CS^{Tm^{\prime}m}(-x)C\right] \nonumber\\
 &&\left.+ {\rm Tr}\left[ \gamma^{\nu}\gamma_5  S_c^{kk^{\prime}}(x)\gamma_5 \gamma^{\nu^{\prime}}CS^{Tjj^{\prime}}(x)C\right] {\rm Tr}\left[\sigma_{\mu^{\prime}\nu^{\prime}}  S_c^{n^{\prime}n}(-x)\sigma_{\mu\nu} CS^{Tm^{\prime}m}(-x)C\right] \right\} \, ,
\end{eqnarray}
where
\begin{eqnarray}
S^{ij}(x)&=& \frac{i\delta_{ij}\!\not\!{x}}{ 2\pi^2x^4}
-\frac{\delta_{ij}m_s}{4\pi^2x^2}-\frac{\delta_{ij}\langle
\bar{s}s\rangle}{12} +\frac{i\delta_{ij}\!\not\!{x}m_s
\langle\bar{s}s\rangle}{48}-\frac{\delta_{ij}x^2\langle \bar{s}g_s\sigma Gs\rangle}{192}+\frac{i\delta_{ij}x^2\!\not\!{x} m_s\langle \bar{s}g_s\sigma
 Gs\rangle }{1152}\nonumber\\
&& -\frac{ig_s G^{a}_{\alpha\beta}t^a_{ij}(\!\not\!{x}
\sigma^{\alpha\beta}+\sigma^{\alpha\beta} \!\not\!{x})}{32\pi^2x^2}  -\frac{\delta_{ij}x^4\langle \bar{s}s \rangle\langle g_s^2 GG\rangle}{27648}-\frac{1}{8}\langle\bar{s}_j\sigma^{\mu\nu}s_i \rangle \sigma_{\mu\nu}    +\cdots \, ,
\end{eqnarray}
\begin{eqnarray}
S_c^{ij}(x)&=&\frac{i}{(2\pi)^4}\int d^4k e^{-ik \cdot x} \left\{
\frac{\delta_{ij}}{\!\not\!{k}-m_c}
-\frac{g_sG^n_{\alpha\beta}t^n_{ij}}{4}\frac{\sigma^{\alpha\beta}(\!\not\!{k}+m_c)+(\!\not\!{k}+m_c)
\sigma^{\alpha\beta}}{(k^2-m_c^2)^2}\right.\nonumber\\
&&\left. -\frac{g_s^2 (t^at^b)_{ij} G^a_{\alpha\beta}G^b_{\mu\nu}(f^{\alpha\beta\mu\nu}+f^{\alpha\mu\beta\nu}+f^{\alpha\mu\nu\beta}) }{4(k^2-m_c^2)^5}+\cdots\right\} \, , \end{eqnarray}
\begin{eqnarray}
f^{\lambda\alpha\beta}&=&(\!\not\!{k}+m_c)\gamma^\lambda(\!\not\!{k}+m_c)\gamma^\alpha(\!\not\!{k}+m_c)\gamma^\beta(\!\not\!{k}+m_c)\, ,\nonumber\\
f^{\alpha\beta\mu\nu}&=&(\!\not\!{k}+m_c)\gamma^\alpha(\!\not\!{k}+m_c)\gamma^\beta(\!\not\!{k}+m_c)\gamma^\mu(\!\not\!{k}+m_c)\gamma^\nu(\!\not\!{k}+m_c)\, ,
\end{eqnarray}
and  $t^n=\frac{\lambda^n}{2}$ \cite{Reinders85}, then compute  the integrals both in the coordinate space and in the momentum space,  and obtain the correlation functions $\Pi_{\mu\mu^\prime}(p)$ (i.e. $\Pi^1_{\mu\mu^\prime}(p)$ and $\Pi^2_{\mu\mu^\prime}(p)$), therefore the QCD spectral densities through dispersion relation $\rho(s)={\rm lim}_{\varepsilon\to 0}\frac{{\rm Im}\Pi(s+i\varepsilon)}{\pi}$. For technical details, one can consult Ref.\cite{WangHuangTao-3900}.

 Now  we  take the
quark-hadron duality below the continuum thresholds $s_0$ and perform Borel transform  with respect to
the variable $P^2=-p^2$ to obtain  the  QCD sum rules:
\begin{eqnarray}
\lambda^2_{X}\, \exp\left(-\frac{M^2_{X}}{T^2}\right)= \int_{4m_c^2}^{s_0} ds\, \rho(s) \, \exp\left(-\frac{s}{T^2}\right) \, ,
\end{eqnarray}
where $\rho(s)=\rho_{TA}(s)$ and $\rho_{TV}(s)$ for the $[sc]_T[\bar{s}\bar{c}]_A+[sc]_V[\bar{s}\bar{c}]_A$ and $[sc]_T[\bar{s}\bar{c}]_V-[sc]_V[\bar{s}\bar{c}]_T$
 axialvector tetraquark states  respectively,
\begin{eqnarray}
\rho_{TA}(s)&=&\rho_{0}(s)+\rho_{3}(s) +\rho_{4}(s)+\rho_{5}(s)+\rho_{6}(s)+\rho_{7}(s) +\rho_{8}(s)+\rho_{10}(s)\, , \nonumber\\
\rho_{TV}(s)&=&\rho_{TA}(s)\mid_{m_c \to -m_c}\, ,
\end{eqnarray}
 the subscripts $i$ in the QCD spectral densities $\rho_{i}(s)$ denote the dimensions of the vacuum condensates,
\begin{eqnarray}
\rho_{3}(s)&\propto& \langle\bar{s}s\rangle\, ,\nonumber\\
\rho_{4}(s)&\propto& \langle\frac{\alpha_{s}GG}{\pi}\rangle\, ,\nonumber\\
\rho_{5}(s)&\propto& \langle\bar{s}g_s\sigma Gs\rangle\, ,\nonumber\\
\rho_{6}(s)&\propto& \langle\bar{s}s\rangle^2\, ,\nonumber\\
\rho_{7}(s)&\propto& \langle\bar{s}s\rangle\langle\frac{\alpha_{s}GG}{\pi}\rangle\, ,\nonumber\\
\rho_{8}(s)&\propto& \langle\bar{s}s\rangle\langle\bar{s}g_s\sigma Gs\rangle\, ,\nonumber\\
\rho_{10}(s)&\propto&  \langle\bar{s}g_s\sigma Gs\rangle^2\, ,\, \langle\bar{s}s\rangle^2\langle\frac{\alpha_{s}GG}{\pi}\rangle\, ,
\end{eqnarray}
the lengthy expressions of the QCD spectral densities are given in Appendix.

  We derive   Eq.(11) with respect to  $\tau=\frac{1}{T^2}$, then eliminate the
 pole residues  $\lambda_{X}$ to obtain the QCD sum rules for the masses,
 \begin{eqnarray}
M_{X}^2&=&- \frac{\int_{4m_c^2}^{s_0} ds \, \frac{d}{d \tau }\,\rho(s)e^{-\tau s}}{\int_{4m_c^2}^{s_0} ds \rho(s)e^{-\tau s}}\, .
\end{eqnarray}

At the QCD side, we take the vacuum condensates  to be the standard values
$\langle\bar{q}q \rangle=-(0.24\pm 0.01\, \rm{GeV})^3$,  $\langle\bar{s}s \rangle=(0.8\pm0.1)\langle\bar{q}q \rangle$,
 $\langle\bar{s}g_s\sigma G s \rangle=m_0^2\langle \bar{s}s \rangle$,
$m_0^2=(0.8 \pm 0.1)\,\rm{GeV}^2$, $\langle \frac{\alpha_s
GG}{\pi}\rangle=(0.33\,\rm{GeV})^4 $    at the energy scale  $\mu=1\, \rm{GeV}$
\cite{SVZ79,Reinders85,ColangeloReview}, and  take the $\overline{MS}$ masses $m_{c}(m_c)=(1.275\pm0.025)\,\rm{GeV}$ and $m_s(\mu=2\,\rm{GeV})=(0.095\pm0.005)\,\rm{GeV}$
 from the Particle Data Group \cite{PDG}.
Moreover,  we take into account
the energy-scale dependence of  the quark condensate, mixed quark condensate and $\overline{MS}$ masses from the renormalization group equation,
 \begin{eqnarray}
 \langle\bar{s}s \rangle(\mu)&=&\langle\bar{s}s \rangle(Q)\left[\frac{\alpha_{s}(Q)}{\alpha_{s}(\mu)}\right]^{\frac{12}{33-2n_f}}\, , \nonumber\\
 \langle\bar{s}g_s \sigma Gs \rangle(\mu)&=&\langle\bar{s}g_s \sigma Gs \rangle(Q)\left[\frac{\alpha_{s}(Q)}{\alpha_{s}(\mu)}\right]^{\frac{2}{33-2n_f}}\, ,\nonumber\\
m_c(\mu)&=&m_c(m_c)\left[\frac{\alpha_{s}(\mu)}{\alpha_{s}(m_c)}\right]^{\frac{12}{33-2n_f}} \, ,\nonumber\\
m_s(\mu)&=&m_s({\rm 2GeV} )\left[\frac{\alpha_{s}(\mu)}{\alpha_{s}({\rm 2GeV})}\right]^{\frac{12}{33-2n_f}}\, ,\nonumber\\
\alpha_s(\mu)&=&\frac{1}{b_0t}\left[1-\frac{b_1}{b_0^2}\frac{\log t}{t} +\frac{b_1^2(\log^2{t}-\log{t}-1)+b_0b_2}{b_0^4t^2}\right]\, ,
\end{eqnarray}
  where $t=\log \frac{\mu^2}{\Lambda^2}$, $b_0=\frac{33-2n_f}{12\pi}$, $b_1=\frac{153-19n_f}{24\pi^2}$, $b_2=\frac{2857-\frac{5033}{9}n_f+\frac{325}{27}n_f^2}{128\pi^3}$,  $\Lambda=210\,\rm{MeV}$, $292\,\rm{MeV}$  and  $332\,\rm{MeV}$ for the flavors  $n_f=5$, $4$ and $3$, respectively \cite{PDG,Narison-mix}, and evolve all the input parameters to the typical energy scales   $\mu$ satisfying the energy scale formula $\mu=\sqrt{M^2_{X/Y/Z}-(2{\mathbb{M}}_c)^2}$ to extract the masses of the axialvector tetraquark states.

We search for the optimal  Borel parameters $T^2$ and threshold parameters $s_0$  to satisfy   the  following four criteria:\\
$\bf 1.$ Pole dominance at the phenomenological side;\\
$\bf 2.$ Convergence of the operator product expansion;\\
$\bf 3.$ Appearance of the Borel platforms;\\
$\bf 4.$ Satisfying the  energy scale formula,\\
  via try and error, and  obtain the Borel windows $T^2$, threshold parameters $s_0$, optimal  energy scales of the QCD spectral densities, and pole contributions of the ground states, see Table 3.

  Now we take a short digression to illustrate how to impose the four criteria to choose the Borel parameters $T^2$ and continuum threshold parameters $s_0$. Firstly, we set $M_X=3.9\,\rm{GeV}$ tentatively, and obtain the energy scale $\mu=1.4\,\rm{GeV}$ according to the  energy scale formula. Then we take the continuum threshold parameters to be  $\sqrt{s_0}=(3.9+0.5)\,\rm{GeV}$ as the energy gap between the ground state and the first radial excited state is about $(0.4-0.6)\,\rm{GeV}$, and obtain the predicted masses $M_X$, pole contributions, and the contributions of the vacuum condensates of dimension $10$. We observe that the predicted masses $M_X$ are much larger than $3.9\,\rm{GeV}$ and the pole contributions are much smaller than $50\%$ in the regions where the Borel platforms appear, furthermore, the contributions of the vacuum condensates of dimension $10$ are not small enough. Then we choose the masses $M_X>3.9\,\rm{GeV}$, say $M_X=4.0\,\rm{GeV}$, $4.1\,\rm{GeV}$, $\cdots$ and reiterate the same procedure  until obtain the optimal   Borel parameters $T^2$ and continuum threshold parameters $s_0$ satisfying the four criteria.

  From Table 3, we can see that the pole dominance  criterion is well satisfied.  In calculations, we observe that the  contributions of the vacuum condensates of dimension  $10$  are $\ll 1\%$ (about $1\%$) in the QCD sum rules for the current $J^1_\mu(x)$($J^2_\mu(x)$), the operator product expansion is well convergent.  We take into account all uncertainties of the input parameters,
and obtain the values of the masses and pole residues of
 the   axialvector tetraquark states, see Table 3 and Figs.1-2. From Table 3, we can see that the energy scale formula is well satisfied. From Figs.1-2 and Table 3, we can see that there appear  platforms in the Borel windows. The four  criteria are all satisfied, our predictions are reliable.

From Table 3, we can see the predicted mass $M_X=4.14\pm0.10\,\rm{GeV}$ for the $[sc]_T[\bar{s}\bar{c}]_V-[sc]_V[\bar{s}\bar{c}]_T$ axialvector tetraquark state is in excellent agreement with the experimental data $4146.5\pm 4.5^{+ 4.6}_{-2.8}\,\rm{MeV}$ from the LHCb collaboration \cite{LHCb-16061,LHCb-16062}, which supports assigning the $X(4140)$ to be the $[sc]_T[\bar{s}\bar{c}]_V-[sc]_V[\bar{s}\bar{c}]_T$  tetraquark state with $J^{PC}=1^{++}$. While the $[sc]_T[\bar{s}\bar{c}]_A+[sc]_V[\bar{s}\bar{c}]_A$ axialvector tetraquark state has a much larger mass than that of the $X(4140)$.

\begin{table}
\begin{center}
\begin{tabular}{|c|c|c|c|c|c|c|c|}\hline\hline
               &$T^2(\rm{GeV}^2)$   &$\sqrt{s_0}(\rm{GeV})$  &$\mu(\rm{GeV})$  &pole          &$M(\rm{GeV})$  &$\lambda(\rm{GeV}^5)$ \\ \hline

$J^1_\mu(x)$   &$4.4-5.0$           &$5.7\pm0.1$             &3.7              &$(40-60)\%$   &$5.20\pm0.11$  &$(2.01\pm0.24)\times10^{-1}$   \\ \hline
$J^2_\mu(x)$   &$2.7-3.3$           &$4.7\pm0.1$             &2.0              &$(41-69)\%$   &$4.14\pm0.10$  &$(4.30\pm0.85)\times10^{-2}$   \\ \hline
\end{tabular}
\end{center}
\caption{ The Borel  windows, continuum threshold parameters, ideal energy scales, pole contributions,   masses and pole residues for the axialvector
  tetraquark states. }
\end{table}

\begin{figure}
 \centering
 \includegraphics[totalheight=5cm,width=7cm]{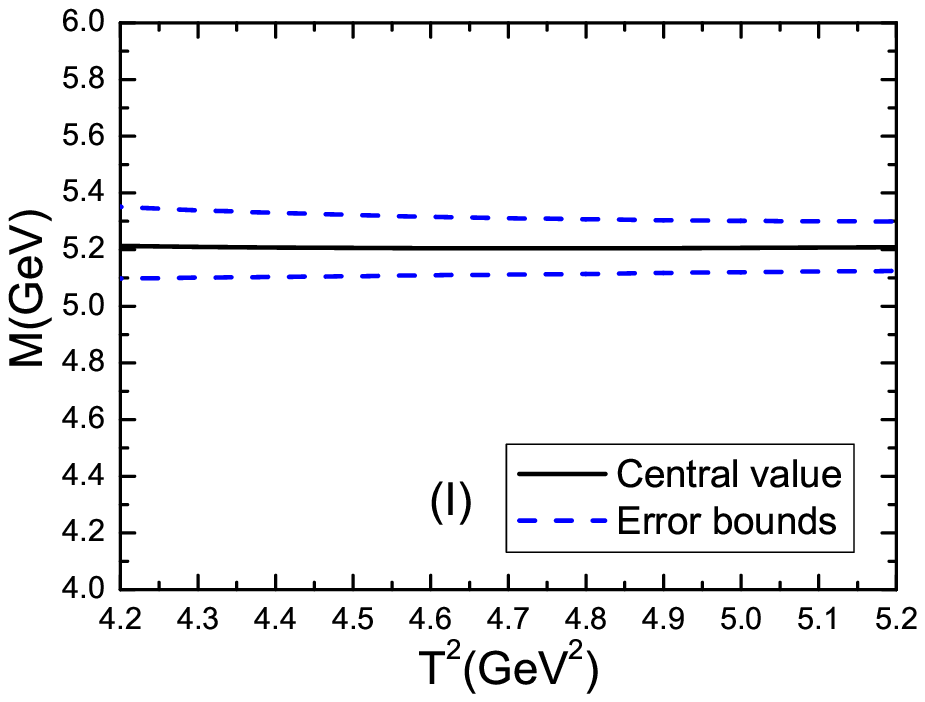}
 \includegraphics[totalheight=5cm,width=7cm]{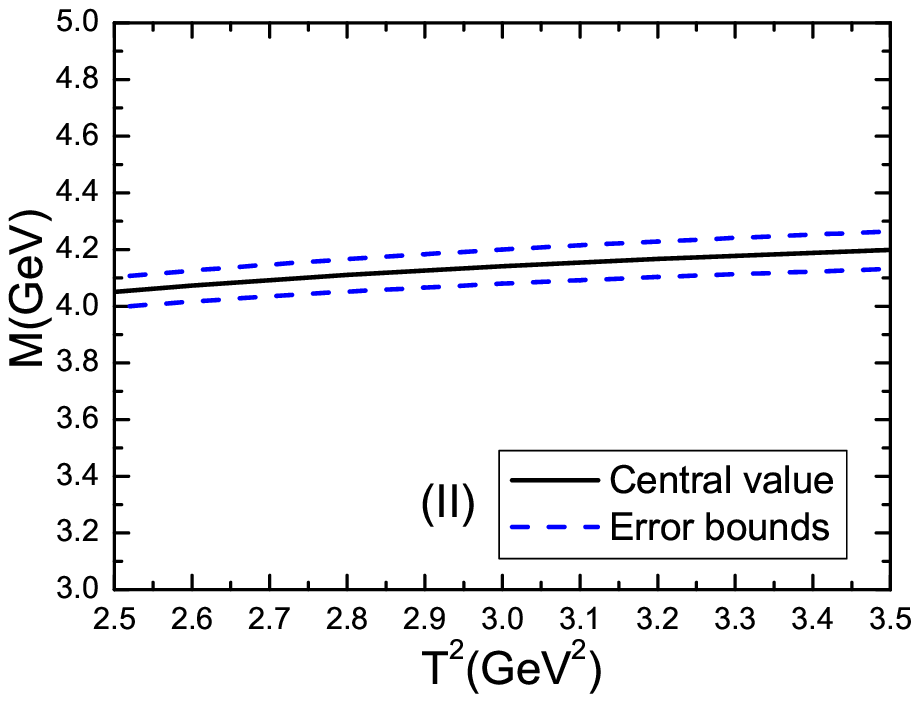}
   \caption{ The masses of the axialvector tetraquark states with variations of the Borel parameters $T^2$, where  the (I) and (II) correspond to the currents $J^1_\mu(x)$ and $J^2_\mu(x)$, respectively.   }
\end{figure}

\begin{figure}
 \centering
 \includegraphics[totalheight=5cm,width=7cm]{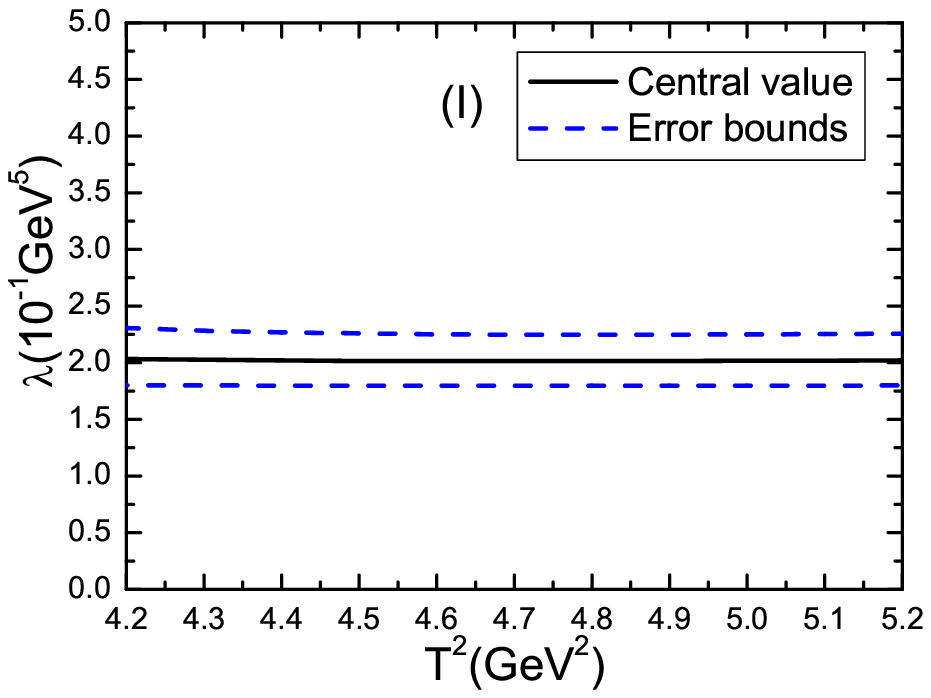}
 \includegraphics[totalheight=5cm,width=7cm]{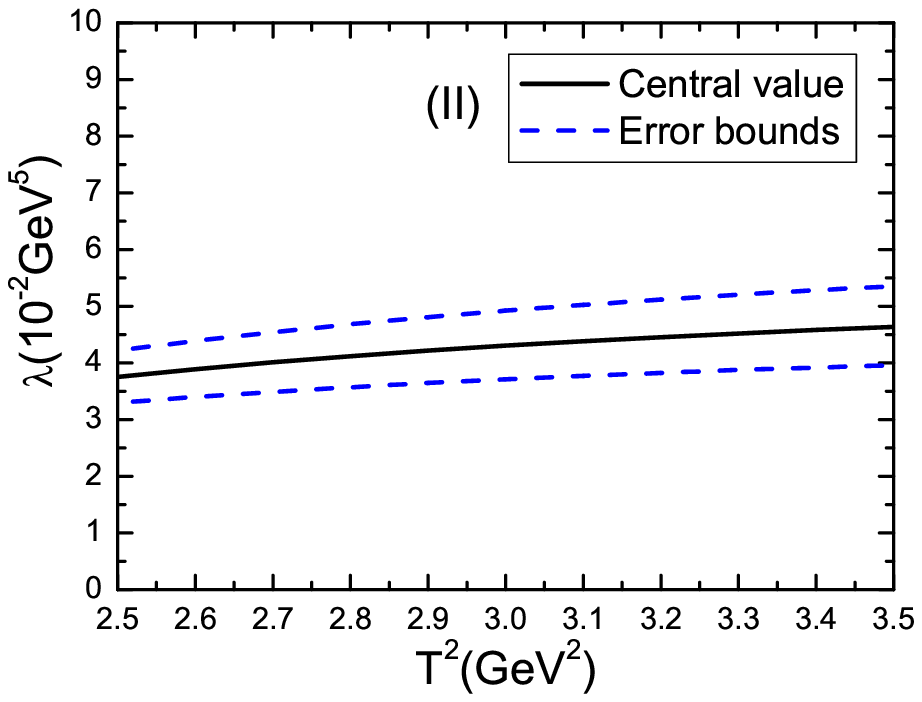}
   \caption{ The pole residues of the axialvector tetraquark states with variations of the Borel parameters $T^2$, where  the (I) and (II) correspond to the currents $J^1_\mu(x)$ and $J^2_\mu(x)$, respectively.   }
\end{figure}

\section{The width of the $X(4140)$ as the axialvector tetraquark state }

We can study the two-body strong decay $X(4140)\to J/\psi \phi$  with the  three-point correlation function
$\Pi_{\alpha\beta\mu}(p,q)$,
\begin{eqnarray}
\Pi_{\alpha\beta\mu}(p,q)&=&i^2\int d^4xd^4y e^{ipx}e^{iqy}\langle 0|T\left\{J_\alpha^{J/\psi}(x)J_{\beta}^{\phi}(y)J_{\mu}^{\dagger}(0)\right\}|0\rangle\, ,
\end{eqnarray}
where the currents
\begin{eqnarray}
J_\alpha^{J/\psi}(x)&=&\bar{c}(x)\gamma_\alpha c(x) \, ,\nonumber \\
J_\beta^{\phi}(y)&=&\bar{s}(y)\gamma_\beta s(y) \, ,\nonumber \\
J_\mu(0)&=&J^2_\mu(0)\, ,
\end{eqnarray}
interpolate the mesons $J/\psi$, $\phi(1020)$ and $X(4140)$ respectively,
\begin{eqnarray}
\langle0|J_{\alpha}^{J/\psi}(0)|J/\psi(p)\rangle&=&f_{J/\psi}m_{J/\psi}\xi_\alpha \,\, , \nonumber \\
\langle0|J_{\beta}^{\phi}(0)|\phi(q)\rangle&=&f_{\phi}m_{\phi}\zeta_\beta \,\, ,
\end{eqnarray}
the $f_{J/\psi}$  and $f_{\phi}$ are the decay constants, the $\xi_\alpha$  and $\zeta_\beta$ are polarization vectors of the mesons $J/\psi$  and $\phi(1020)$, respectively. In this section, we will use the notation $m_X$ in stead of $M_X$ for the special case $J_\mu(0)=J^2_\mu(0)$.

At the phenomenological side,  we insert  a complete set of intermediate hadronic states with
the same quantum numbers as the current operators $J_\alpha^{J/\psi}(x)$, $J_\beta^{\phi}(y)$, $J_{\mu}^{\dagger}(0)$ into the three-point
correlation function $\Pi_{\alpha\beta\mu}(p,q)$ and  isolate the ground state
contributions to obtain the  result,
\begin{eqnarray}
\Pi_{\alpha\beta\mu}(p,q)&=&  \frac{f_{\phi}m_{\phi} f_{J/\psi}m_{J/\psi}\lambda_{X}g_{XJ/\psi \phi} }{(m_{X}^2-p^{\prime2})(m_{J/\psi}^2-p^2)(m_{\phi}^2-q^2)} \,\varepsilon^{\lambda\tau\rho\theta}p^{\prime}_\lambda\left(-g_{\mu\tau}+\frac{p_{\mu}^{\prime}p^{\prime}_{\tau}}{p^{\prime 2}} \right) \left(-g_{\alpha\rho}+\frac{p_{\alpha}p_{\rho}}{p^{ 2}} \right)\nonumber\\
&&\left(-g_{\beta\theta}+\frac{q_{\beta}q_{\theta}}{q^{ 2}} \right)+\cdots  \nonumber\\
&=&\left\{ \frac{f_{\phi}m_{\phi} f_{J/\psi}m_{J/\psi}\lambda_{X}g_{XJ/\psi \phi} }{(m_{X}^2-p^{\prime2})(m_{J/\psi}^2-p^2)(m_{\phi}^2-q^2)} + \frac{1}{(m_{X}^2-p^{\prime2})(m_{J/\psi}^2-p^2)} \int_{s^0_\phi}^\infty dt\frac{\rho_{X\phi^\prime}(p^{\prime 2},p^2,t)}{t-q^2}\right.\nonumber\\
&&\left.+ \frac{1}{(m_{X}^2-p^{\prime2})(m_{\phi}^2-q^2)} \int_{s^0_{J/\psi}}^\infty dt\frac{\rho_{X\psi^\prime}(p^{\prime 2},t,q^2)}{t-p^2}\right. \nonumber\\
&&\left.+ \frac{1}{(m_{J/\psi}^2-p^{2})(m_{\phi}^2-q^2)} \int_{s^0_{X}}^\infty dt\frac{\rho_{X^{\prime}J/\psi}(t,p^2,q^2)+\rho_{X^{\prime}\phi}(t,p^2,q^2)}{t-p^{\prime2}}+\cdots\right\}\nonumber\\
&&\left(\varepsilon_{\alpha\beta\mu\lambda}p^\lambda+\cdots\right) +\cdots\, ,\nonumber\\
&=&\Pi(p^{\prime2},p^2,q^2) \, \varepsilon_{\alpha\beta\mu\lambda}p^\lambda+\cdots
\end{eqnarray}
where $p^\prime=p+q$,  the $g_{XJ/\psi\phi}$  is the hadronic coupling constant defined by
\begin{eqnarray}
\langle J/\psi(p,\xi)\phi(q,\zeta)|X(p^{\prime},\varepsilon)\rangle&=&i g_{XJ/\psi\phi} \, \varepsilon^{\lambda\tau\rho\theta} p^\prime_\lambda \varepsilon_\tau \xi^*_\rho \zeta^*_\theta\, ,
\end{eqnarray}
the four   functions $\rho_{X\phi^\prime}(p^{\prime 2},p^2,t)$, $ \rho_{X\psi^\prime}(p^{\prime 2},t,q^2)$,
$ \rho_{X^{\prime}J/\psi}(t^\prime,p^2,q^2)$ and $\rho_{X^{\prime}\phi}(t^\prime,p^2,q^2)$
   have complex dependence on the transitions between the ground states and the higher resonances  or the continuum states.

In this article, we choose the tensor structure $\varepsilon_{\alpha\beta\mu\lambda}p^\lambda$  to study the  $g_{XJ/\psi\phi}$   to avoid the contaminations from the relevant  scalar and pseudoscalar mesons according to the non-vanishing coupling constants,
\begin{eqnarray}
\langle0|J_{\alpha}^{J/\psi}(0)|\chi_{c0}(p)\rangle&=&f_{\chi_{c0}}p_\alpha \,\, , \nonumber \\
\langle0|J_{\beta}^{\phi}(0)|f_0(q)\rangle&=&f_{f_0}q_\beta \,\, , \nonumber \\
\langle X_{0}(p^\prime)|J_\mu^{\dagger}(0)|0\rangle&=&-i\,\lambda_{X_{0}}\,p^\prime_\mu\,\, ,
\end{eqnarray}
where the $f_{\chi_{c0}}$, $f_{f_0}$ and $\lambda_{X_{0}}$ are the decay constants of the $\chi_{c0}(3414)$, $f_0(980)$   and $X_{0}(J^P=0^-)$, respectively.

We introduce the parameters $C_{X\phi^\prime}$, $C_{X\psi^\prime}$, $C_{X^\prime\phi}$ and $C_{X^\prime J/\psi}$   to parameterize the net effects,
\begin{eqnarray}
C_{X\phi^\prime}&=&\int_{s^0_\phi}^\infty dt\frac{ \rho_{X\phi^\prime}(p^{\prime 2},p^2,t)}{t-q^2}\, ,\nonumber\\
C_{X\psi^\prime}&=&\int_{s^0_{J/\psi}}^\infty dt\frac{\rho_{X\psi^\prime}(p^{\prime 2},t,q^2)}{t-p^2}\, ,\nonumber\\
C_{X^\prime\phi}&=&\int_{s^0_{X}}^\infty dt\frac{ \rho_{X^\prime\phi}(t,p^2,q^2)}{t-p^{\prime2}}\, ,\nonumber\\
C_{X^\prime J/\psi}&=&\int_{s^0_{X}}^\infty dt\frac{ \rho_{X^\prime J/\psi}(t,p^2,q^2)}{t-p^{\prime2}}\, .
\end{eqnarray}
Then the correlation function $\Pi(p^{\prime2},p^2,q^2)$ on the phenomenological side can be written as
\begin{eqnarray}
\Pi(p^{\prime2},p^2,q^2)&=& \frac{f_{\phi}m_{\phi} f_{J/\psi}m_{J/\psi}\lambda_{X}g_{XJ/\psi \phi} }{(m_{X}^2-p^{\prime2})(m_{J/\psi}^2-p^2)(m_{\phi}^2-q^2)} + \frac{C_{X\phi}}{(m_{X}^2-p^{\prime2})(m_{J/\psi}^2-p^2)}  \nonumber\\
&&+ \frac{C_{XJ/\psi}}{(m_{X}^2-p^{\prime2})(m_{\phi}^2-q^2)}  +\frac{C_{X^{\prime}J/\psi}+C_{X^{\prime}\phi}}{(m_{J/\psi}^2-p^{2})(m_{\phi}^2-q^2)}  +\cdots\, .
\end{eqnarray}

Now we carry out the operator product expansion up to the vacuum condensates of dimension 5 and neglect the tiny contributions of the gluon condensate.
 The correlation function $\Pi_{QCD}(p^{\prime 2},p^2,q^2)$ can be written as
\begin{eqnarray}
\Pi_{QCD}(p^{\prime 2},p^2,q^2)&=&  \int_{4m_c^2}^{s^0_{J/\psi}}ds \int_0^{u^0_{\phi}}du  \frac{\rho_{QCD}(p^{\prime2},s,u)}{(s-p^2)(u-q^2)}+\cdots\, ,
\end{eqnarray}
through dispersion relation, where the $\rho_{QCD}(p^{\prime 2},s,u)$   is the QCD spectral density,
\begin{eqnarray}
\rho_{QCD}(p^{\prime 2},s,u)&=& {\lim_{\epsilon_2\to 0}} \,\,{\lim_{\epsilon_1\to 0}}\,\,\frac{  {\rm Im}_{s}\,{\rm Im}_{u}\,\Pi_{QCD}(p^{\prime 2},s+i\epsilon_2,u+i\epsilon_1) }{\pi^2} \, ,
\end{eqnarray}
we introduce the subscript $QCD$ to denote the $QCD$ side.

We rewrite the correlation function $\Pi_H(p^{\prime 2},p^2,q^2)$ on the hadron  side as
\begin{eqnarray}
\Pi_{H}(p^{\prime 2},p^2,q^2)&=&\int_{(m_{J/\psi}+m_{\phi})^2}^{s_{X}^0}ds^\prime \int_{4m_c^2}^{s^0_{J/\psi}}ds \int_0^{u^0_{\phi}}du  \frac{\rho_H(s^\prime,s,u)}{(s^\prime-p^{\prime2})(s-p^2)(u-q^2)}+\cdots\, ,
\end{eqnarray}
 through dispersion relation, where the $\rho_H(s^\prime,s,u)$   is the hadronic spectral density,
\begin{eqnarray}
\rho_H(s^\prime,s,u)&=&{\lim_{\epsilon_3\to 0}}\,\,{\lim_{\epsilon_2\to 0}} \,\,{\lim_{\epsilon_1\to 0}}\,\,\frac{ {\rm Im}_{s^\prime}\, {\rm Im}_{s}\,{\rm Im}_{u}\,\Pi_H(s^\prime+i\epsilon_3,s+i\epsilon_2,u+i\epsilon_1) }{\pi^3} \, ,
\end{eqnarray}
we introduce the subscript $H$ to denote the hadron side.
However, on the QCD side, the QCD spectral density $\rho_{QCD}(s^\prime,s,u)$ does  not exist,
\begin{eqnarray}
\rho_{QCD}(s^\prime,s,u)&=&{\lim_{\epsilon_3\to 0}}\,\,{\lim_{\epsilon_2\to 0}} \,\,{\lim_{\epsilon_1\to 0}}\,\,\frac{ {\rm Im}_{s^\prime}\, {\rm Im}_{s}\,{\rm Im}_{u}\,\Pi_{QCD}(s^\prime+i\epsilon_3,s+i\epsilon_2,u+i\epsilon_1) }{\pi^3} \nonumber\\
&=&0\, ,
\end{eqnarray}
because
\begin{eqnarray}
{\lim_{\epsilon_3\to 0}}\,\,\frac{ {\rm Im}_{s^\prime}\,\Pi_{QCD}(s^\prime+i\epsilon_3,p^2,q^2) }{\pi} &=&0\, .
\end{eqnarray}

We math the hadron side of the correlation function  with the QCD side of the correlation function,
and carry out the integral over $ds^\prime$  firstly to obtain the solid duality \cite{WangZhang-Solid},
\begin{eqnarray}
\int_{\Delta_s^2}^{s^0} ds \int_{\Delta_u^2}^{u^0} du \frac{\rho_{QCD}(p^{\prime2},s,u)}{(s-p^2)(u-q^2)}&=&\int_{\Delta_s^2}^{s^0} ds \int_{\Delta_u^2}^{u^0} du \frac{1}{(s-p^2)(u-q^2)}\left[ \int_{\Delta^2}^{\infty} ds^\prime \frac{\rho_{H}(s^{\prime},s,u)}{s^\prime-p^{\prime2}}\right]\, , \nonumber\\
\end{eqnarray}
 the $\Delta_s^2$ and $\Delta_u^2$ denote the thresholds $4m_c^2$ and $0$, the $\Delta^2$ denotes the threshold $(m_{J/\psi}+m_{\phi})^2$.
 Now we write the quark-hadron duality explicitly,
 \begin{eqnarray}
  \int_{4m_c^2}^{s^0_{J/\psi}}ds \int_0^{u^0_{\phi}}du  \frac{\rho_{QCD}(p^{\prime2},s,u)}{(s-p^2)(u-q^2)}&=& \int_{4m_c^2}^{s^0_{J/\psi}}ds \int_0^{u^0_{\phi}}du  \int_{(m_{J/\psi}+m_{\phi})^2}^{\infty}ds^\prime \frac{\rho_H(s^\prime,s,u)}{(s^\prime-p^{\prime2})(s-p^2)(u-q^2)} \nonumber\\
  &=&\frac{f_{\phi}m_{\phi} f_{J/\psi}m_{J/\psi}\lambda_{X}g_{XJ/\psi \phi} }{(m_{X}^2-p^{\prime2})(m_{J/\psi}^2-p^2)(m_{\phi}^2-q^2)}  +\frac{C_{X^{\prime}J/\psi}+C_{X^{\prime}\phi}}{(m_{J/\psi}^2-p^{2})(m_{\phi}^2-q^2)} \, . \nonumber\\
\end{eqnarray}
 No approximation is needed, we do not need the continuum threshold parameter $s^0_{X}$ in the $s^\prime$ channel. The present approach was introduced in Ref.\cite{WangZhang-Solid}.

In numerical calculations,   we   take the functions $C_{X\phi^\prime}$, $C_{X\psi^\prime}$, $C_{X^\prime\phi}$ and $C_{X^\prime J/\psi}$  as free parameters, and choose the suitable values  to eliminate the contaminations from the higher resonances and continuum states to obtain the stable QCD sum rules with the variations of
the Borel parameters.
We  set $p^{\prime2}=p^2$  and perform the double Borel transform  with respect to the variables $P^2=-p^2$ and $Q^2=-q^2$, respectively  to obtain the  QCD sum rules,
\begin{eqnarray}
&& \frac{f_{\phi}m_{\phi} f_{J/\psi}m_{J/\psi}\lambda_{X}g_{XJ/\psi \phi}}{m_{X}^2-m_{J/\psi}^2} \left[ \exp\left(-\frac{m_{J/\psi}^2}{T_1^2} \right)-\exp\left(-\frac{m_{X}^2}{T_1^2} \right)\right]\exp\left(-\frac{m_{\phi}^2}{T_2^2} \right) \nonumber\\
&&+\left(C_{X^{\prime}J/\psi}+C_{X^{\prime}\phi}\right) \exp\left(-\frac{m_{J/\psi}^2}{T_1^2} -\frac{m_{\phi}^2}{T_2^2} \right)\nonumber\\
&&=-\frac{1}{16\sqrt{2}\pi^4}\int_{4m_c^2}^{s^0_{J/\psi}} ds \int_{0}^{s^0_{\phi}} du  u\sqrt{1-\frac{4m_c^2}{s}}\left(m_c-\frac{m_s}{2}-\frac{m_s m_c^2}{s} \right)\exp\left(-\frac{s}{T_1^2} -\frac{u}{T_2^2} \right)\nonumber\\
&&+\frac{m_s m_c\langle\bar{s}s\rangle}{2\sqrt{2}\pi^2} \int_{4m_c^2}^{s^0_{J/\psi}} ds \sqrt{1-\frac{4m_c^2}{s}}\exp\left(-\frac{s}{T_1^2}  \right) \nonumber\\
&&-\frac{\langle\bar{s}g_s\sigma Gs\rangle}{36\sqrt{2}\pi^2} \int_{4m_c^2}^{s^0_{J/\psi}} ds \sqrt{1-\frac{4m_c^2}{s}}\frac{s+2m_c^2}{s}\exp\left(-\frac{s}{T_1^2}  \right) \nonumber\\
&&-\frac{m_s m_c\langle\bar{s}g_s\sigma Gs\rangle}{24\sqrt{2}\pi^2 T_2^2} \int_{4m_c^2}^{s^0_{J/\psi}} ds \sqrt{1-\frac{4m_c^2}{s}}\exp\left(-\frac{s}{T_1^2}  \right) \nonumber\\
&&-\frac{m_s m_c\langle\bar{s}g_s\sigma Gs\rangle}{16\sqrt{2}\pi^2 } \int_{4m_c^2}^{s^0_{J/\psi}} ds \frac{1}{\sqrt{s^2-4sm_c^2}}\exp\left(-\frac{s}{T_1^2}  \right) \, .
\end{eqnarray}
In calculations, we observe that there appears divergence due to the endpoint $s=4m_c^2$, we can avoid the endpoint divergence with the simple replacement
 $\frac{1}{\sqrt{s^2-4sm_c^2}} \to \frac{1}{\sqrt{s^2-4sm_c^2+4m_s^2{\rm GeV}^2}}$ by adding a small squared $s$-quark mass $4m_s^2$.

The hadronic parameters are taken as    $m_{\phi}=1.019461\,\rm{GeV}$,
$m_{J/\psi}=3.0969\,\rm{GeV}$ \cite{PDG},
$f_{J/\psi}=0.418 \,\rm{GeV}$  \cite{Becirevic}, $f_{\phi}=0.253\,\rm{GeV}$, $\sqrt{s^0_{\phi}}=1.5\,\rm{GeV}$   \cite{Wang-Y4274}, $\sqrt{s^0_{J/\psi}}=3.6\,\rm{GeV}$,
$M_X=4146.5\,\rm{MeV}$ \cite{LHCb-16061,LHCb-16062},   $\lambda_{X}=4.30\times 10^{-2}\,\rm{GeV}^5$. At the QCD side, we take the energy scale of the QCD spectral density to be $\mu=2\,\rm{GeV}$, just like in the two-point QCD sum rules. Then we set the Borel parameters to be $T_1^2=T_2^2=T^2$ for simplicity.
The unknown parameters are chosen as $C_{X^{\prime}J/\psi}+C_{X^{\prime}\phi}=-0.00261\,\rm{GeV}^7 $   to obtain  platform in the Borel window $T^2=(3.6-4.6)\,\rm{GeV}^2$.

In Fig.3, we plot the hadronic coupling constant  $g_{XJ/\psi \phi}$  with variation of the  Borel parameter $T^2$. From the figure, we can see that there appears platform in the Borel window indeed. After taking into account the uncertainties of the input parameters, we obtain the  hadronic coupling constant   $g_{XJ/\psi \phi}$, \begin{eqnarray}
g_{XJ/\psi \phi} &=&-(2.23\pm0.29)\, .
\end{eqnarray}

Now it is easy to obtain the decay width,
\begin{eqnarray}
\Gamma(X(4140)\to J/\psi \phi)&=& \frac{p\left(m_{X},m_{J/\psi},m_{\phi}\right)}{24\pi m_{X}^2}g_{XJ/\psi\phi}^2\left\{\frac{\left(m_{X}^2-m_{\phi}^2\right)^2}{2m_{J/\psi}^2}+\frac{\left(m_{X}^2-m_{J/\psi}^2\right)^2}{2m_{\phi}^2} \right.\nonumber\\
&&\left. +4m_X^2-\frac{m_{J/\psi}^2+m_\phi^2}{2}\right\} \nonumber\\
&=&86.9\pm22.6\,{\rm{MeV}}\, ,
\end{eqnarray}
where $p(a,b,c)=\frac{\sqrt{[a^2-(b+c)^2][a^2-(b-c)^2]}}{2a}$.	The width $\Gamma(X(4140)\to J/\psi \phi)=86.9\pm22.6\,{\rm{MeV}}$ is in excellent agreement with the experimental data $83\pm 21^{+21}_{-14} {\mbox{ MeV}}$ from the LHCb collaboration \cite{LHCb-16061,LHCb-16062}.
The present work supports assigning the $X(4140)$ to be the  $[sc]_T[\bar{s}\bar{c}]_V-[sc]_V[\bar{s}\bar{c}]_T$  tetraquark state with $J^{PC}=1^{++}$.

\begin{figure}
 \centering
 \includegraphics[totalheight=7cm,width=10cm]{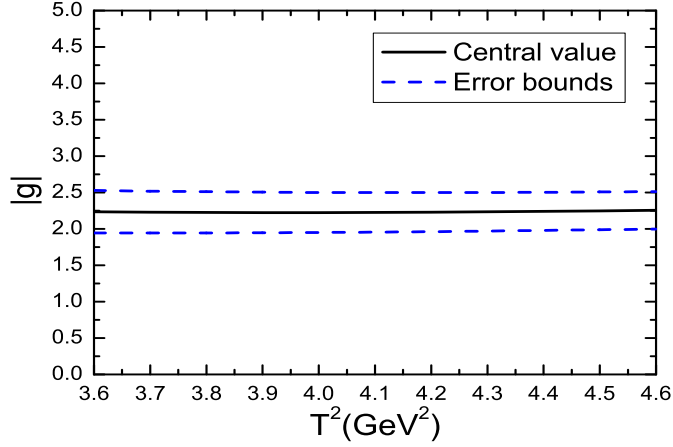}
   \caption{ The hadronic coupling constant $g_{XJ/\psi\phi}$ with variation of the Borel parameter $T^2$.   }
\end{figure}

\section{Conclusion}
In this article, we construct  both the $[sc]_T[\bar{s}\bar{c}]_A+[sc]_A[\bar{s}\bar{c}]_T$ type and $[sc]_T[\bar{s}\bar{c}]_V-[sc]_V[\bar{s}\bar{c}]_T$ type axialvector currents with $J^{PC}=1^{++}$ to study the mass of the $X(4140)$ with the QCD sum rules by carrying out the operator product expansion up to the vacuum condensates of dimension 10 and take the energy scale formula $\mu=\sqrt{M^2_{X/Y/Z}-(2{\mathbb{M}}_c)^2}$ to determine the ideal energy scales of the QCD spectral densities. The predicted masses support assigning the $X(4140)$ to be the $[sc]_T[\bar{s}\bar{c}]_V-[sc]_V[\bar{s}\bar{c}]_T$ type axialvector tetraquark state. Then we calculate the hadronic coupling constant $g_{XJ/\psi\phi}$ with the QCD sum rules based on the solid quark-hadron duality, and obtain the decay width $\Gamma(X(4140)\to J/\psi \phi)=86.9\pm22.6\,{\rm{MeV}}$, which is in excellent agreement with the experimental data $83\pm 21^{+21}_{-14} {\mbox{ MeV}}$ from the LHCb collaboration.  In summary, the present work supports assigning the $X(4140)$ to be the  $[sc]_T[\bar{s}\bar{c}]_V-[sc]_V[\bar{s}\bar{c}]_T$  type tetraquark state with $J^{PC}=1^{++}$.

\section*{Appendix}

The explicit expressions of the QCD spectral densities $\rho_0(s)$, $\rho_3(s)$, $\rho_5(s)$, $\rho_6(s)$, $\rho_7(s)$, $\rho_8(s)$ and  $\rho_{10}(s)$,

\begin{eqnarray}
\rho_0(s)&=&\frac{3 }{1024\pi^6}\int dydz\, yz\left(1-y-z\right)^2\, \left(s-\overline{m}_c^2\right)^3\left(5s-\overline{m}_c^2\right) \nonumber\\
&&-\frac{m_c^2}{768\pi^6}\int dydz\,\left(5+y+z\right)\left(1-y-z\right)^2\left(s-\overline{m}_c^2\right)^3\nonumber\\
&&-\frac{3m_s m_c}{256\pi^6}\int dydz\, y\left(1-y-z\right)^2\left(s-\overline{m}_c^2\right)^2\left(3s-\overline{m}_c^2\right) \, ,
\end{eqnarray}

\begin{eqnarray}
\rho_3(s)&=&\frac{ m_c\langle\bar{s}s\rangle}{32\pi^4} \int dydz\, y\left(1-y-z\right) \left(s-\overline{m}_c^2\right)\left(7s-3\overline{m}_c^2\right) \nonumber\\
&&+\frac{m_s\langle\bar{s}s\rangle}{32\pi^4} \int dydz\, yz \left(s-\overline{m}_c^2\right)\left(23s-9\overline{m}_c^2\right) \nonumber\\
&&-\frac{m_s m_c^2\langle\bar{s}s\rangle}{16\pi^4} \int dydz\, \left(7+y+z\right)\left(s-\overline{m}_c^2\right) \, ,
\end{eqnarray}

\begin{eqnarray}
\rho_{4}(s)&=&\frac{m_c^2}{1152\pi^4}\langle\frac{\alpha_{s}GG}{\pi}\rangle \int dydz\, \frac{ \left[y^2 + \left(z - 1\right) y - 9 z\right]  \left(1-y-z\right)^2}{y^2}\left(2s-\overline{m}_c^2\right)   \nonumber\\
&&+\frac{m_c^2}{96\pi^4}\langle\frac{\alpha_{s}GG}{\pi}\rangle \int dydz\,
\frac{\left(1-y-z\right)^2}{y^2}\left[\frac{y s}{2}-\left(1-y\right)\left(s-\overline{m}_c^2\right)\right]   \nonumber\\
&&+\frac{m_c^2}{576\pi^4}\langle\frac{\alpha_{s}GG}{\pi}\rangle \int dydz \, \frac{\left(1-y-z\right)^3\left(1-y\right)}{y^2} \left(s-\overline{m}_c^2\right)  \nonumber\\
&&+\frac{m_s m_c^3}{256\pi^4}\langle\frac{\alpha_{s}GG}{\pi}\rangle \int dydz\, \frac{\left(1-y-z\right)^2}{y^2}
\left[1+\frac{2s}{3}\delta\left(s-\overline{m}_c^2\right)\right]  \nonumber\\
&&+\frac{m_s m_c}{128\pi^4}\langle\frac{\alpha_{s}GG}{\pi}\rangle \int dydz\, \frac{z\left(1-y-z\right)^2}{y^2}
\left[\left(\frac{3y}{2}-1\right)\left(s-\overline{m}_c^2\right)-\frac{2s}{3}\right.\nonumber\\
&&\left.+\frac{11ys}{6}+\frac{ys^2}{3}\delta\left(s-\overline{m}_c^2\right)\right]   \nonumber\\
&&+\frac{1}{1536\pi^4}\langle\frac{\alpha_{s}GG}{\pi}\rangle \int dydz\, \left[\left(1-y-z\right)^2+2yz\right]\,s\left(s-\overline{m}_c^2\right) \nonumber\\
&&+\frac{m_c^2}{3072\pi^4}\langle\frac{\alpha_{s}GG}{\pi}\rangle \int dydz\, \frac{\left(6y+3z+9\right)\left(1-y-z\right)}{y}
\left(s-\overline{m}_c^2\right) \nonumber\\
&&+\frac{m_s m_c}{512\pi^4}\langle\frac{\alpha_{s}GG}{\pi}\rangle \int dydz\, \frac{\left(2z-3y\right)\left(1-y-z\right)}{y}\left(5s-3\overline{m}_c^2\right)  \nonumber\\
&&+\frac{m_s m_c}{3072\pi^4}\langle\frac{\alpha_{s}GG}{\pi}\rangle \int dydz\, \frac{\left(1-y-z\right)^2}{z}\left(5s-3\overline{m}_c^2\right) \nonumber\\
&&-\frac{m_c^2}{512\pi^4}\langle\frac{\alpha_{s}GG}{\pi}\rangle \int dydz\, \left(s-\overline{m}_c^2\right) \nonumber\\
&&-\frac{7m_c^2}{3072\pi^4}\langle\frac{\alpha_{s}GG}{\pi}\rangle \int dydz\, \frac{\left(1-y-z\right)^2}{yz}\left(s-\overline{m}_c^2\right) \nonumber\\
&&+\frac{13m_c^2}{18432\pi^4}\langle\frac{\alpha_{s}GG}{\pi}\rangle \int dydz\, \frac{\left(1-y-z\right)^3}{yz}\left(s-\overline{m}_c^2\right)\, ,
\end{eqnarray}

\begin{eqnarray}
\rho_5(s)&=&-\frac{m_c \langle\bar{s}g_{s}\sigma Gs\rangle}{64\pi^4} \int dydz\, y\left(5s-3\overline{m}_c^2\right) \nonumber\\
&&-\frac{m_s\langle\bar{s}g_{s}\sigma Gs\rangle}{32\pi^4} \int dy\, y\left(1-y\right)\left(7s-4\widetilde{m}_c^2\right) \nonumber\\
&&+\frac{11m_s m_c^2\langle\bar{s}g_{s}\sigma Gs\rangle}{96\pi^4} \int dy
-\frac{m_s m_c^2\langle\bar{s}g_{s}\sigma Gs\rangle}{96\pi^4} \int dydz \nonumber\\
&&+\frac{m_c \langle\bar{s}g_{s}\sigma Gs\rangle}{768\pi^4}  \int dydz\, \frac{ 5y^2 - 3\left(z + 2\right) y - 9\left(z - 1\right)z}{y}\left(5s-3\overline{m}_c^2\right) \nonumber\\
&&-\frac{m_s \langle\bar{s}g_{s}\sigma Gs\rangle}{256\pi^4}  \int dydz\, z\, \left(5s-3\overline{m}_c^2\right) \, ,
\end{eqnarray}

\begin{eqnarray}
\rho_6(s)&=&-\frac{\langle\bar{s}s\rangle^2}{12\pi^2} \int dy\, y\left(1-y\right)\left(5s-3\widetilde{m}_c^2\right)
+ \frac{m_c^2\langle\bar{s}s\rangle^2}{4\pi^2}  \int dy \nonumber\\
&&+\frac{m_s m_c\langle\bar{s}s\rangle^2}{8\pi^2}  \int dy\, y\left[1+\frac{2s}{3}\delta\left(s-\widetilde{m}_c^2\right)\right]\, ,
\end{eqnarray}

\begin{eqnarray}
\rho_{7}(s)&=&-\frac{m_c^3\langle\bar{s}s\rangle}{288\pi^2}\langle\frac{\alpha_{s}GG}{\pi}\rangle \int dydz\,
\frac{\left(1-y-z\right)}{y^2}\left(1+\frac{2s}{T^2}\right)\delta\left(s-\overline{m}_c^2\right)   \nonumber\\
&&+\frac{m_c\langle\bar{s}s\rangle}{16\pi^2}\langle\frac{\alpha_{s}GG}{\pi}\rangle \int dydz\,
\frac{z\left(1-y-z\right)}{y^2}\left[\frac{1}{3}-\frac{y}{2}+\left(\frac{2s}{9}-\frac{7ys}{18}-\frac{ys^2}{9T^2}\right)
\delta\left(s-\overline{m}_c^2\right)\right]   \nonumber\\
&&-\frac{m_s m_c^2\langle\bar{s}s\rangle}{72\pi^2}\langle\frac{\alpha_{s}GG}{\pi}\rangle \int dydz\,
\frac{z}{y^2}\left(1+\frac{7s}{2T^2}\right)\delta\left(s-\overline{m}_c^2\right)   \nonumber\\
&&-\frac{m_s m_c^2\langle\bar{s}s\rangle}{18\pi^2}\langle\frac{\alpha_{s}GG}{\pi}\rangle \int dydz\,
\frac{1}{y^2}\left(2-y-\frac{y s}{T^2}\right)\delta\left(s-\overline{m}_c^2\right)   \nonumber\\
&&+\frac{m_s m_c^2\langle\bar{s}s\rangle}{72\pi^2}\langle\frac{\alpha_{s}GG}{\pi}\rangle \int dydz\,
\frac{\left(1-y-z\right)\left(1-y\right)}{y^2}\delta\left(s-\overline{m}_c^2\right)   \nonumber\\
&&-\frac{m_s m_c^2\langle\bar{s}s\rangle}{144\pi^2T^2}\langle\frac{\alpha_{s}GG}{\pi}\rangle \int dydz\, \frac{\left(1-y-z\right)}{y}
\,s\,\delta\left(s-\overline{m}_c^2\right)   \nonumber\\
&&+\frac{m_c\langle\bar{s}s\rangle}{128\pi^2} \langle\frac{\alpha_{s}GG}{\pi}\rangle \int dydz\,
\frac{3y-2z}{y}\left[1+\frac{2s}{3}\delta\left(s-\overline{m}_c^2\right) \right]   \nonumber\\
&&+\frac{m_s\langle\bar{s}s\rangle}{768\pi^2} \langle\frac{\alpha_{s}GG}{\pi}\rangle \int dy\,s\, \delta\left(s-\widetilde{m}_c^2\right)  \nonumber\\
&&-\frac{m_s m_c^2\langle\bar{s}s\rangle}{768\pi^2} \langle\frac{\alpha_{s}GG}{\pi}\rangle \int dydz\, \frac{1}{y} \delta\left(s-\overline{m}_c^2\right)  \nonumber\\
&&-\frac{m_c\langle\bar{s}s\rangle}{384\pi^2}\langle\frac{\alpha_{s}GG}{\pi}\rangle \int dydz\,
\frac{\left(1-y-z\right)}{z}\left[1+\frac{2s}{3}\delta\left(s-\overline{m}_c^2\right)\right]\nonumber\\
&&+\frac{m_s\langle\bar{s}s\rangle}{128\pi^2}\langle\frac{\alpha_{s}GG}{\pi}\rangle \int dydz\, \left[1+\frac{8s}{9}\delta\left(s-\overline{m}_c^2\right)\right] \nonumber\\
&&-\frac{m_s m_c^2\langle\bar{s}s\rangle}{2304\pi^2}\langle\frac{\alpha_{s}GG}{\pi}\rangle \int dydz\, \frac{25+13y+13z}{yz} \delta\left(s-\overline{m}_c^2\right) \nonumber\\
&&+\frac{m_c \langle\bar{s}s\rangle}{192\pi^2} \langle\frac{\alpha_{s}GG}{\pi}\rangle  \int dy\, y\left[1+\frac{2s}{3}\delta\left(s-\widetilde{m}_c^2\right)\right] \nonumber\\
&&+\frac{m_s \langle\bar{s}s\rangle}{32\pi^2} \langle\frac{\alpha_{s}GG}{\pi}\rangle  \int dy\,
y\left(1-y\right)\left[1+\left(\frac{7s}{9}+\frac{2s^2}{9T^2}\right)\delta\left(s-\widetilde{m}_c^2\right)\right] \nonumber\\
&&-\frac{m_s m_c^2 \langle\bar{s}s\rangle}{96\pi^2} \langle\frac{\alpha_{s}GG}{\pi}\rangle  \int dy\,
\left(1+\frac{s}{T^2}\right)\delta\left(s-\widetilde{m}_c^2\right) \, ,
\end{eqnarray}

\begin{eqnarray}
\rho_8(s)&=&\frac{\langle\bar{s}s\rangle \langle\bar{s}g_{s}\sigma Gs\rangle}{8\pi^2} \int dy\, y\left(1-y\right)
\left[3+\left(\frac{7s}{3}+\frac{2s^2}{3 T^2}\right)\delta\left(s-\widetilde{m}_c^2\right)\right]\nonumber\\
&&-\frac{m_c^2\langle\bar{s}s\rangle \langle\bar{s}g_{s}\sigma Gs\rangle}{8\pi^2} \int dy\,\left(1+\frac{s}{T^2}\right)\delta\left(s-\widetilde{m}_c^2\right)
\nonumber\\
&&-\frac{5m_s m_c\langle\bar{s}s\rangle \langle\bar{s}g_{s}\sigma Gs\rangle}{144\pi^2} \int dy\, y
\left(1+\frac{3s}{2T^2}+\frac{s^2}{T^4}\right)\delta\left(s-\widetilde{m}_c^2\right)\nonumber\\
&&+\frac{\langle\bar{s}s\rangle \langle\bar{s}g_{s}\sigma Gs\rangle}{128\pi^2} \int dy\,
\left[1+\frac{2s}{3}\delta\left(s-\widetilde{m}_c^2\right)\right]\nonumber\\
&&+\frac{m_s m_c\langle\bar{s}s\rangle \langle\bar{s}g_{s}\sigma Gs\rangle}{128\pi^2} \int dy\,
\frac{1-y}{y}\left(1+\frac{2s}{T^2}\right)\delta\left(s-\widetilde{m}_c^2\right) \nonumber\\
&&-\frac{m_s m_c\langle\bar{s}s\rangle \langle\bar{s}g_{s}\sigma Gs\rangle}{192\pi^2} \int dy\,
\left(1+\frac{2s}{T^2}\right)\delta\left(s-\widetilde{m}_c^2\right) \, ,
\end{eqnarray}

\begin{eqnarray}
\rho_{10}(s)&=&-\frac{\langle\bar{s}g_{s}\sigma Gs\rangle^2}{32\pi^2} \int dy\, y\left(1-y\right)
\left\{1+\frac{4s}{3 T^2}+\frac{5s^2}{6T^4}-\frac{s^3}{6T^6}\right)\delta\left(s-\widetilde{m}_c^2\right)\nonumber\\
&&-\frac{m_s m_c\langle\bar{s}g_{s}\sigma Gs\rangle^2}{288\pi^2T^2} \int dy\, y \left(1+\frac{s}{T^2}+\frac{s^2}{2T^4}-\frac{s^3}{T^6}
\right)\delta\left(s-\widetilde{m}_c^2\right) \nonumber\\
&&-\frac{m_c^2\langle\bar{s}s\rangle^2}{108T^2} \langle\frac{\alpha_{s}GG}{\pi}\rangle \int dy\,
\frac{\left(1-y\right)}{y^2}\left(1-\frac{2s}{T^2}\right)\delta\left(s-\widetilde{m}_c^2\right)  \nonumber\\
&&-\frac{m_c^2\langle\bar{s}s\rangle^2}{18T^2} \langle\frac{\alpha_{s}GG}{\pi}\rangle \int dy\,\frac{1}{y^2}\left(1-\frac{y s}{2T^2}\right)\delta\left(s-\widetilde{m}_c^2\right)  \nonumber\\
&&+\frac{m_s m_c^3\langle\bar{s}s\rangle^2}{144T^4}\langle\frac{\alpha_{s}GG}{\pi}\rangle \int dy\,
\frac{1}{y^2}\left(1-\frac{2s}{3T^2}\right)\delta\left(s-\widetilde{m}_c^2\right)  \nonumber\\
&&+\frac{m_s m_c\langle\bar{s}s\rangle^2}{216T^2}\langle\frac{\alpha_{s}GG}{\pi}\rangle \int dy\,
\frac{1-y}{y^2}\left(\frac{y}{2}-1+\frac{2s}{T^2}+\frac{y s}{2T^2}
-\frac{y s^2}{T^4}\right)\delta\left(s-\widetilde{m}_c^2\right) \nonumber\\
&&+\frac{m_s m_c\langle\bar{s}s\rangle^2}{1728T^2}\langle\frac{\alpha_{s}GG}{\pi}\rangle \int dy\,
\frac{1}{1-y}\left(1-\frac{2s}{T^2}\right)\delta\left(s-\widetilde{m}_c^2\right)\nonumber\\
&&-\frac{\langle\bar{s}s\rangle^2}{576}\langle\frac{\alpha_{s}GG}{\pi}\rangle \int dy\,
\left(1-\frac{2s}{T^2}\right)\delta\left(s-\widetilde{m}_c^2\right)\nonumber\\
&&-\frac{\langle\bar{s}g_{s}\sigma Gs\rangle^2}{768\pi^2} \int dy\, \left(1+\frac{3s}{2T^2}+\frac{s^2}{T^4}\right)\delta\left(s-\widetilde{m}_c^2\right) \nonumber\\
&&+\frac{m_s m_c\langle\bar{s}g_{s}\sigma Gs\rangle^2}{768\pi^2T^2} \int dy\,
\frac{1-y}{y}\left(1+\frac{s}{T^2}-\frac{2s^2}{T^4}\right)\delta\left(s-\widetilde{m}_c^2\right) \nonumber\\
&&-\frac{m_s m_c\langle\bar{s}g_{s}\sigma Gs\rangle^2}{1152\pi^2T^2} \int dy\,
\left(1+\frac{s}{T^2}-\frac{2s^2}{T^4}\right)\delta\left(s-\widetilde{m}_c^2\right) \nonumber\\
&&+\frac{7\langle\bar{s}g_{s}\sigma Gs\rangle^2}{27648\pi^2} \int dy\,\left(1+\frac{2s}{ T^2}\right)\delta\left(s-\widetilde{m}_c^2\right)\nonumber\\
&&-\frac{\langle\bar{s}s\rangle^2}{36} \langle\frac{\alpha_{s}GG}{\pi}\rangle \int dy\,
 y\left(1-y\right) \left(1+\frac{4s}{3T^2}+\frac{5s^2}{6T^4}-\frac{s^3}{6T^6}\right)\delta\left(s-\widetilde{m}_c^2\right) \nonumber\\
&&-\frac{m_s m_c\langle\bar{s}s\rangle^2}{432T^2} \langle\frac{\alpha_{s}GG}{\pi}\rangle \int dy\,
 y \left(1+\frac{s}{T^2}+\frac{s^2}{2T^4}-\frac{s^3}{T^6}\right)\delta\left(s-\widetilde{m}_c^2\right) \, ,
\end{eqnarray}
 where $\int dydz=\int_{y_i}^{y_f}dy \int_{z_i}^{1-y}dz$, $y_{f}=\frac{1+\sqrt{1-4m_c^2/s}}{2}$,
$y_{i}=\frac{1-\sqrt{1-4m_c^2/s}}{2}$, $z_{i}=\frac{y
m_c^2}{y s -m_c^2}$, $\overline{m}_c^2=\frac{(y+z)m_c^2}{yz}$,
$ \widetilde{m}_c^2=\frac{m_c^2}{y(1-y)}$, $\int_{y_i}^{y_f}dy \to \int_{0}^{1}dy$, $\int_{z_i}^{1-y}dz \to \int_{0}^{1-y}dz$,  when the $\delta$ functions $\delta\left(s-\overline{m}_c^2\right)$ and $\delta\left(s-\widetilde{m}_c^2\right)$ appear.

\section*{Acknowledgements}
This  work is supported by National Natural Science Foundation, Grant Number  11775079.


\begin{thebibliography}{99}

\bibitem{CDF0903}  T. Aaltonen et al,   Phys. Rev. Lett. {\bf 102} (2009) 242002.


\bibitem{CDF1101} T. Aaltonen et al,  Mod. Phys. Lett. {\bf A32} (2017) 1750139.

\bibitem{CMS1309} S. Chatrchyan et al, Phys. Lett. {\bf B734} (2014) 261.

\bibitem{D0-1309}  V. M. Abazov et al,  Phys. Rev. {\bf D89} (2014) 012004.

\bibitem{D0-1508}  V. M. Abazov et al, Phys. Rev. Lett. {\bf 115} (2015) 232001.

\bibitem{LHCb-16061}  R. Aaij et al,  Phys. Rev. Lett. {\bf 118} (2017) 022003.

\bibitem{LHCb-16062}  R. Aaij et al,  Phys. Rev. {\bf D95} (2017) 012002.


\bibitem{PDG}  M. Tanabashi et al, Phys. Rev. {\bf  D98} (2018)  030001.

\bibitem{X4140-tetraquark-Stancu}   F. Stancu, J. Phys. {\bf G37} (2010) 075017.



\bibitem{Wang1502-Y4140}  Z. G. Wang and  Y. F. Tian, Int. J. Mod. Phys. {\bf A30} (2015) 1550004.

\bibitem{X4140-tetraquark-Lebed} R. F. Lebed and A. D. Polosa,  Phys. Rev. {\bf D93} (2016) 094024.

\bibitem{Maiani-X4140} L. Maiani, A. D. Polosa and V. Riquer, Phys. Rev. {\bf D94} (2016)  054026.

\bibitem{X4140-tetraquark} R. Zhu, Phys. Rev. {\bf D94} (2016)  054009;
Q. Fang Lu and Y. B. Dong, Phys. Rev. {\bf D94} (2016)  074007;
J. Wu, Y. R. Liu, K. Chen, X. Liu and S. L. Zhu, Phys. Rev. {\bf D94} (2016)  094031;
M. N. Anwar, J. Ferretti and E. Santopinto, Phys. Rev. {\bf D98} (2018) 094015.


\bibitem{X4140-hybrid} N. Mahajan,  Phys. Lett. {\bf B679} (2009) 228.

\bibitem{X4140-hybrid-Wang0903}  Z. G. Wang,  Eur. Phys. J. {\bf C63} (2009) 115;
Z. G. Wang, Z. C. Liu and X. H. Zhang,  Eur. Phys. J. {\bf C64} (2009) 373.


\bibitem{X4140-rescat} X. H. Liu, Phys. Lett. {\bf B766} (2017) 117.



\bibitem{Wang1606-Y3915C}  Z. G. Wang, Eur. Phys. J. {\bf C77} (2017) 78.



\bibitem{Wang1607-Y3915A} Z. G. Wang, Eur. Phys. J. {\bf A53} (2017) 19.

\bibitem{Wang1607-Y4140} Z. G. Wang, Eur. Phys. J. {\bf C76} (2016) 657.


\bibitem{ChenZhu-2011} W. Chen and S. L. Zhu, Phys. Rev. {\bf D83} (2011) 034010.

\bibitem{Chen1606} H. X. Chen, E. L. Cui, W. Chen, X. Liu and S. L. Zhu, Eur. Phys. J. {\bf C77} (2017) 160.

\bibitem{Azizi1703} S. S. Agaev, K. Azizi and H. Sundu, Phys. Rev. {\bf 95} (2017)  114003.

\bibitem{ChenZhu1706} W. Chen, H. X. Chen, X. Liu, T. G. Steele and S. L. Zhu, Phys. Rev. {\bf D96} (2017) 114017.

\bibitem{Wang-tetra-formula}  Z. G. Wang, Eur. Phys. J. {\bf C74} (2014)  2874; Z. G. Wang and T. Huang, Nucl. Phys. {\bf A930} (2014) 63.

\bibitem{WangHuang-molecule} Z. G. Wang  and T. Huang, Eur. Phys. J. {\bf C74} (2014)  2891;  Z. G. Wang, Eur. Phys. J. {\bf C74} (2014)  2963.


\bibitem{Wang1508-EPJC}
Z. G. Wang, Eur. Phys. J. {\bf C76} (2016) 70;
Z. G. Wang  and T. Huang, Eur. Phys. J. {\bf C76} (2016)  43;
Z. G. Wang, Nucl. Phys. {\bf B913} (2016) 163.


\bibitem{WangEPJC-1601} Z. G. Wang, Eur. Phys. J. {\bf C76} (2016)  387.

\bibitem{SVZ79} M. A. Shifman, A. I. Vainshtein and V. I. Zakharov, Nucl. Phys. {\bf B147} (1979) 385; Nucl. Phys. {\bf B147} (1979) 448.

\bibitem{Reinders85} L. J. Reinders, H. Rubinstein and S. Yazaki, Phys. Rept. {\bf 127} (1985) 1.

\bibitem{WangHuangTao-3900} Z. G. Wang and T. Huang,  Phys. Rev. {\bf D89} (2014) 054019.


\bibitem{ColangeloReview} P. Colangelo and A. Khodjamirian, hep-ph/0010175.


\bibitem{Narison-mix} S. Narison and R. Tarrach, Phys. Lett. {\bf 125 B} (1983) 217.


\bibitem{WangZhang-Solid} Z. G. Wang and  J. X. Zhang, Eur. Phys. J. {\bf C78} (2018) 14.

\bibitem{Becirevic} D. Becirevic, G. Duplancic, B. Klajn, B. Melic and F. Sanfilippo,  Nucl. Phys. {\bf B883} (2014) 306.

\bibitem{Wang-Y4274}  Z. G. Wang, Eur. Phys. J. {\bf C77} (2017)  174.


\end{thebibliography}
\end{document}